# "This Browser is Lightning Fast": The Effects of Message Content on Perceived Performance

**Jess Hohenstein[1], Bill Selman[2], Gemma Petrie[3], Jofish Kaye[4], & Rebecca Weiss[5]**
Mozilla
Mountain View, CA, USA
jch378@cornell.edu[1], wselman@gmail.com[2], gpetrie@mozilla.com[3], jofish@jofish.com[4],
rweiss@mozilla.com[5]

**ABSTRACT**
With technical performance being similar for various web browsers, improving user perceived performance is integral to optimizing browser quality. We investigated the importance of priming, which has a well-documented ability to affect people's beliefs, on users' perceptions of web browser performance. We studied 1495 participants who read either an article about performance improvements to Mozilla Firefox, an article about user interface updates to Firefox, or an article about self-driving cars, and then watched video clips of browser tasks. As the priming effect would suggest, we found that reading articles about Firefox increased participants' perceived performance of Firefox over the most widely used web browser, Google Chrome. In addition, we found that article content mattered, as the article about performance improvements led to higher performance ratings than the article about UI updates. Our findings demonstrate how perceived performance can be improved without making technical improvements and that designers and developers must consider a wider picture when trying to improve user attitudes about technology.

**ACM Classification Keywords**
H.5.2. User Interfaces: Theory and Methods

**INTRODUCTION**
Despite the wide variety of apps that provide a great deal of Internet-based functionality, the general-purpose web browser remains prevalent in daily life on both desktop computers and mobile devices. Although there are numerous web browsers available, Google Chrome is the most frequently used [47]. From a technical standpoint, the prevalence of Chrome usage may seem surprising, given that the performance of various browsers is rated as similar by industry reviewers [22]. Coupled with the fact that time-related factors (e.g. response time) have been identified as some of the key factors important to user satisfaction with software [34], one might expect web browser usage distribution to be more even. So why is one browser experiencing much wider usage than others? Our research seeks to investigate how other factors affect user evaluation of web browsers.

Perceived performance refers to how quickly software appears to perform a given task and is an integral element of building user trust and holding attention [4, 6]. Specifically, web browser performance specs are a primary metric deployed in software benchmarks, technology media news stories, and related marketing copy for comparing web browsers. With technical performance being very similar across various web browsers, investigating other factors that contribute to user perceived performance is integral to understanding and improving holistic perceptions of browser quality.

A variety of factors affect user perceived performance. Previous research on search engine results and social media has illustrated the importance of popularity and brand on usage and perception of performance, respectively [17, 28]. Additionally, multiple studies [15, 38] have illustrated the effect that priming users with information from credible sources can have on their opinions. This paper builds on these findings through a large-scale survey investigating whether priming can be used to improve users' perceived performance of a less well-known browser over the most widely-used web browser. In doing so, this paper makes several contributions:

- A demonstration of the power of priming in the HCI domain and its importance as a factor in user perceived performance
- An example of how designers and developers can improve software perceived performance without making technical improvements by distributing specifically-curated content

**BACKGROUND**
The idea of priming is based on psychological research showing that people rarely take into consideration all available relevant information when making decisions. To minimize cognitive effort, people satisfice rather than optimize [36, 37], drawing on a limited subset of information instead of going through a comprehensive process of reviewing and interpreting all available information. This subset of information often consists of whatever is most accessible and comes to the mind quickly [13, 42].

The priming effect of the media on people's attitudes and behaviors is a widely-studied phenomenon [2, 16, 21].

According to priming theory, the most accessible information will be used for decision-making, so topics presented prevalently in the media will likely be readily available in consumers' minds and could alter their judgments and opinions. This occurrence has been well-documented in the literature, with a multitude of examples of priming being illustrated in the media [8, 19, 26, 28, 35, 39], with longitudinal analyses of survey data matching the content analysis of media coverage [3], as well as in consumer product evaluations [5, 23, 24, 43].

Given what we know about priming, it seems possible that priming could be utilized to improve perceptions of different browsers. In other words, does priming offer a potential option for newer or smaller companies to improve perceived performance of their software? Furthermore, does the content of the priming message matter? Priming is often thought of as an extension of agenda setting, which refers to the idea that the emphasis placed on a subject is related to its importance among mass audiences [20]. Therefore, if we want to rule out the possibility that we are simply agenda setting by repeatedly mentioning a specific browser, we must investigate whether reading performance-related content about the browser increases the likelihood of producing a priming effect compared to reading generic content about the browser.

## THE CURRENT STUDY

Despite similarities in technical performance [30], Google Chrome is much more widely used than other web browsers [41]. Given the widely-reported influence of branding, media, and marketing on beliefs and attitudes [21, 25, 26] taken with Google's dominance in branding [9], the media [33], and advertising [38], including its primary marketing message that Chrome is faster than other browsers [12, 29], its widespread usage over other browsers is not surprising. How can less popular web browsers without the resources of a multinational conglomerate compete? We believe that previous research presents an opportunity to investigate the ability to contend with a dominant brand by priming users with messages about another brand to change their perceptions regarding performance. Our hypotheses are as follows:

Hypothesis 1A: **Chrome is seen as faster by default**. Users who do not receive a priming message about Firefox will rate Chrome as faster than Firefox.

Hypothesis 1B: **Firefox is seen as faster after reading about Firefox UI**. Priming users with a message about Firefox UI improvements will lead to higher perceived performance ratings for Firefox than for Chrome.

Hypothesis 1C: **Firefox is seen as faster after reading about Firefox performance**. Priming users with a message about the improved performance of Firefox will lead to higher perceived performance ratings for Firefox than for Chrome.

Hypothesis 1D: **Firefox is seen as faster after reading about performance improvements compared to UI improvements**. The message about improved performance of Firefox will lead to higher ratings of perceived performance for Firefox than the message about the Firefox UI.

If true, these hypotheses would illustrate the importance of publicly distributing curated messages in order to affect the technology perceptions of users, which would have wide-ranging implications for future research, user experience testing, and product development as a whole.

## METHOD

To investigate our hypotheses, we studied participants' perceived performance of Mozilla Firefox and Google Chrome with and without reading a browser-related article beforehand. We used a between-subjects design and manipulated the article content (Self-Driving Cars, Firefox UI, Firefox Performance) and article source (USA Today, The Verge). Each participant experienced one of the six combinations of article content and source. A breakdown of our design and the number of participants in each branch can be seen in Table 1.

| Branch | Source | Article | N |
|---|---|---|---|
| Control | USA Today | Self-Driving Cars | 235 |
| | The Verge | Self-Driving Cars | 251 |
| Exp. Group 1 | USA Today | Firefox UI | 246 |
| | The Verge | Firefox UI | 250 |
| Exp. Group 2 | USA Today | Firefox Performance | 273 |
| | The Verge | Firefox Performance | 240 |
| | | Total | 1495 |

**Table 1: A breakdown of our study design conditions and the number of participants in each branch.**

### Materials

*Articles*

Participants read one of three articles:

*Self-Driving Cars*: "Korea is building a 'city' for self-driving cars" served as the control article. The article details the opening of a large testing facility for self-driving cars that includes most things a car might encounter in a real driving situation. While still falling under the umbrella of tech, the article was unrelated to any variables of interest in this study. The article contained 180 words and 1,143 characters and did not contain the word "Firefox".

*Firefox UI*: "Latest Firefox release includes bold and modern 'Photon' interface updates" served as our first treatment article. The article details new UI updates to Firefox that make the browser more usable and modern than its competitors. The language used in the article was carefully curated to not include any words that could be related to speed or performance. The article contains 165 words, 1,112 characters, and contained the word "Firefox" 6 times.

*Firefox Performance*: "Project Quantum makes latest Firefox release lightning fast" served as our second treatment article. The article specifically details aspects of the new Firefox browser that make it faster, smoother, and higher-performing than competitors. The article contains 188 words, 1,242 characters, and contained the word "Firefox" 5 times.

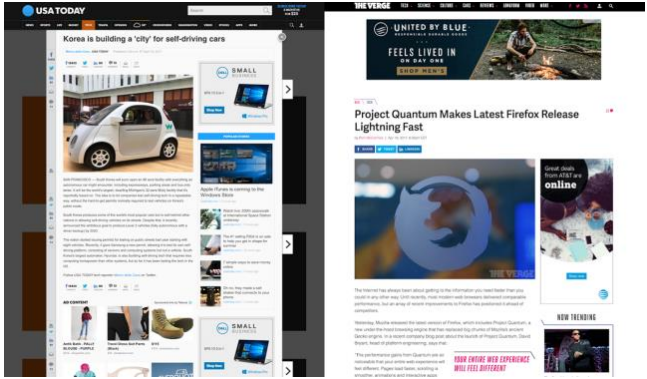

**Figure 1: Each participant read one of 6 articles, as detailed in Table 1. Shown above are the USA Today x Self-Driving Cars (left) and The Verge x Firefox Performance (right) articles.**

Two versions of each article were created to appear to come from different sources: one from USA Today, a general national news source, and one from The Verge, a tech news source. This gave us six conditions in total. With these two sources, we targeted differing levels of credibility. To further increase perception of credibility in accordance with previous research [7], statements in the article were supported by objective, valid, and strong arguments. All three articles presented contained approximately the same number of words and were equally cognitively demanding.

*Browser Interaction Videos*
All participants viewed four short (3-second duration) video clips of typical browser interactions that they rated in terms of performance. The first set of clips showed Mozilla Firefox and Google Chrome opening a new tab, and the second set of clips showed Mozilla Firefox and Google Chrome opening a new window. Participants were able to replay the videos as many times as desired. The videos of each process were the same duration for each browser.

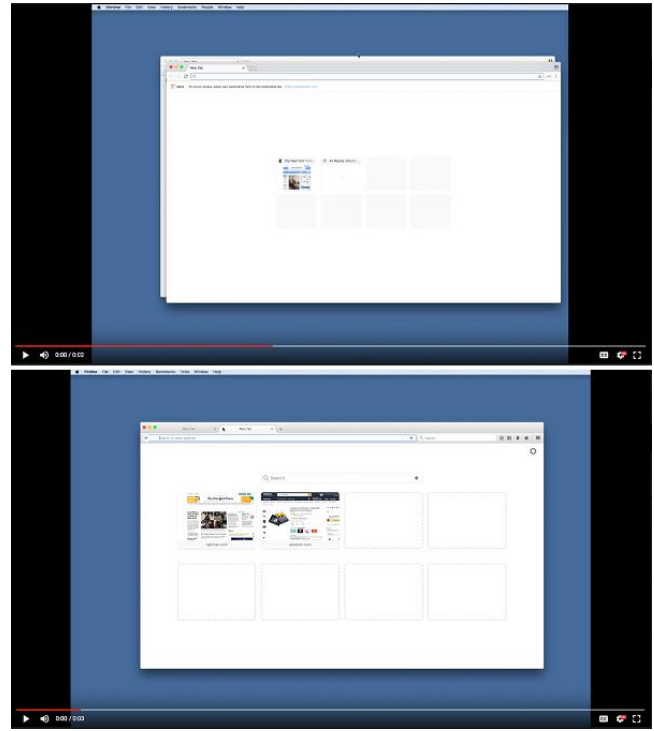

**Figure 2: Participants each viewed 4 videos: Mozilla Chrome opening a new window and a new tab and Google Chrome opening a new window and a new tab. Shown above are Google Chrome opening a new window (top) and Mozilla Firefox opening a new tab (bottom).**

**Participants**
Survey Sampling International (SSI), a web-based survey panel system, was used to enlist 1495 participants (793 female, 702 male), who received gift cards or monetary compensation for their participation. 535 additional participants were not included in the analysis due to partial completion of the survey. Demographic screening criteria based on Pew's omnibus Internet survey [10] were used to ensure a diverse, demographically representative sample of United States Internet users. Participants were 53% female, average age 45.8 (SD=16.6); 64.4% white, 12.1% Hispanic/LatinX, 12.1% Black/African American, 8.5% Asian; 41.8% college graduates; median household income $35,000-$49,999.

**Procedure**
Participants first answered questions about their technological experience and behavior. These questions covered topics including giving advice to/installing software for friends and family, average internet use, primarily-used internet browser, and frequency of getting news online. Next, participants randomly read one of the three articles from one of the two sources. Participants then rated the credibility of the article source using a 7-item Likert-type credibility scale adapted from Gaziano and McGrath [11].

Next, participants were presented with a task that involved giving feedback about two browsers, Google Chrome and

Mozilla Firefox. They viewed videos of Google Chrome opening a new window, Mozilla Firefox opening a new window, Google Chrome opening a new tab, and Mozilla Firefox opening a new tab. The videos were all equal in duration, but asking intermittent questions allowed us to conceal this. After viewing each video, participants rated their agreement with statements regarding speed and preference (e.g. "The wait for the page to load was: Extremely slow - Extremely fast"; "The opening of the page was: Extremely unimaginative - Extremely innovative"; "The loading of the page was: Extremely uneven - Extremely smooth") using 7-point rating scales. After answering questions about the individual videos, participants were asked forced comparison questions about both browsers (e.g. "Which browser opened the tab more quickly?"; "Which browser was the smoothest when opening the new tab?"; "Which browser had a more innovative tab opening?") with multiple choice options of Google Chrome, Mozilla Firefox, or Neither. We asked questions about factors other than speed to conceal the fact that perception of speed was our main concern. To minimize order bias or recency effects (where a message presented last has greater influence on the decision than messages received earlier [14]), half of participants saw the tab opening videos first, and the other half saw the window opening videos first.

Next, participants were asked further comparison questions about the two browsers (e.g. "Overall, [browser] was: Extremely slow - Extremely fast"; "Overall, [browser] was: Extremely uneven - Extremely smooth") using 7-point rating scales. Participants also answered two multiple choice questions regarding which browser was the fastest and smoothest overall, with options of Google Chrome, Mozilla Firefox, and Neither.

Last, participants answered demographic questions including age, gender, ethnicity, education, and income.

## RESULTS

In undertaking our quantitative comparison analysis of survey responses, we treated the Self-Driving Cars article and USA Today source as control conditions to compare against the treatment conditions. Results were analyzed and will be discussed as follows. First, to determine any main treatment effects, we calculated the proportion of endorsements of Firefox or Chrome (or neither) as faster against participants' preferred browser and ran Chi-squared tests for independence for hypothesis testing. Second, as Chi-squared tests only allow for significance testing of the marginal distribution of responses, we sought to estimate individual-level effects through examining the likelihood of endorsing Firefox as the faster browser as a result of treatment. To determine the likelihood of the priming effect, we assigned a dummy variable for each respondent indicating whether they rated Firefox as faster. We then fit logit models predicting whether Firefox was considered faster against the treatment conditions (i.e. the article content and article source that the respondent saw), as well as other variables such as the respondent's preferred browser. Significance is at the α=0.05 level unless otherwise noted.

|  | N | % |
|---|---|---|
| Google Chrome | 189 | 38.65 |
| Mozilla Firefox | 151 | 30.88 |
| Neither | 149 | 30.47 |

Table 3: Participants in the control condition who endorsed Chrome, Firefox, or Neither as faster. More participants rated Chrome as faster (38.65%) than Firefox (30.88%).

**Chrome is Fastest by Default**
In the control condition, more participants rated Google Chrome as faster than Mozilla Firefox (38.65% rated Chrome as faster; 30.88% rated Firefox as faster; 30.47% rated neither as faster, $\chi(1)=6.23$, $p=0.04$), as shown in Table 3. This result supports our hypothesis that Chrome is perceived as faster by default. The preference for endorsing neither as faster over Firefox could also suggest that Firefox is not well-known among the general United States population.

| Preferred browser | Browser endorsed as faster | | |
|---|---|---|---|
|  | Google Chrome | Mozilla Firefox | Neither |
| Mozilla Firefox | 21.4% | 51% | 27.6% |
| Google Chrome | 48.5% | 23.9% | 27.6% |
| Other | 30.3% | 30.3% | 39.5% |
| **Total** | 38.7% | 30.9% | 30.5% |

$\chi^2=37.948 \cdot df=4 \cdot \varphi=0.279 \cdot p=0.000$

Table 4: Participants who endorsed Chrome, Firefox, or Neither as faster by self-reported browser preference. Strong brand preferences are illustrated within preferred browser groups, with 51% of Firefox-preferring respondents endorsing Firefox as faster and 48.5% of Chrome-preferring respondents endorsing Chrome as faster.

We also examined the proportion of endorsements for the faster browser within each group of preferred browsers (recoded to Firefox, Chrome, or any other browser), reported in Table 4. We observed strong brand preferences within self-reported preferred browser groups. 48.5% of respondents who preferred Chrome endorsed Chrome as faster, whereas 51% of respondents who preferred Firefox rated Firefox as faster. Similarly, 39.5% of alternative browser users endorsed neither as faster ($\chi(4)=37.948$, p=<0.001).

**Priming About Firefox Makes Firefox Faster**
When respondents read one of the priming articles, we saw a prevalence of the priming effect over brand preferences,

providing support for the hypothesis that Firefox is faster after reading about Firefox improvements. First, we examined the endorsements against preferred browsers within the performance treatment condition, in order to examine the influence of the content treatment in isolation. These results, summarized in Table 5, show that 56.6% of respondents who preferred Firefox, 54.2% of respondents who preferred alternative browsers, and 43.8% of respondents who preferred Chrome endorsed Firefox over Chrome at the α=0.1 level ($\chi(4)$=9.054, p=0.06), providing weak evidence that priming effects can overcome brand preferences. This result shows that the effect of viewing priming messages about performance may be somewhat resilient against brand preferences.

| Preferred browser | Browser endorsed as faster | | |
|---|---|---|---|
| | Google Chrome | Mozilla Firefox | Neither |
| Mozilla Firefox | 14.5% | 56.6% | 28.9% |
| Google Chrome | 27.4% | 43.8% | 28.8% |
| Other | 21.4% | 54.2% | 24.4% |
| **Total** | 23.8% | 48.5% | 27.7% |

$\chi^2=9.054 \cdot df=4 \cdot \varphi=0.133 \cdot p=0.060$

**Table 5: Participants who read the Firefox Performance content and endorsed Chrome, Firefox, or Neither as faster. Overall, 48.5% (p=.06) of participants endorsed Firefox as fastest, providing weak evidence that priming effects are stronger than brand preferences.**

Supporting the hypothesis that reading about performance is more effective than reading about UI improvements, we also found that the priming message regarding the Firefox UI did not persuade respondents who preferred Chrome as much as the Firefox improved performance message, suggesting that the content of the prime does indeed matter and we are not simply agenda setting, as a prime regarding general improvement was less effective. As shown in Table 6, only 32.5% of Chrome users endorsed Firefox as the faster browser in the UI treatment condition, compared to 48.9% of respondents who preferred Firefox and 45.9% of respondents who preferred alternative browsers ($\chi(4)$=16.298, p=0.003). These results provide partial support for the idea that Firefox is faster after reading about Firefox UI improvements, in that priming messages related to UI do increase the perception of performance, but some brands may attenuate the impact of the prime over others.

**Priming About Performance Makes Firefox Faster Than Chrome**

Next, we review the results of the logit models, which allow for likelihood estimates of the impact of treatment on endorsing Firefox as faster. First, we fit a logit model predicting the endorsement of Firefox as faster against browser preferences with Firefox preference as the base condition. Mirroring the Chi-squared independence tests, if brand preferences are strong, we should observe a lower likelihood of endorsing Firefox as faster given brand preferences. This was the case, as reported in Table 7. We observed that respondents who prefer Chrome were 18% less likely (β=-0.18, p=<.001) to endorse Firefox over all other treatment conditions. In line with brand preferences, participants who preferred other browsers were also significantly less likely to endorse Firefox as the faster browser over all treatment conditions (β=-0.08, p=.034). These results again confirm the hypothesis that Chrome is faster by default.

| Preferred browser | Browser endorsed as faster | | |
|---|---|---|---|
| | Google Chrome | Mozilla Firefox | Neither |
| Mozilla Firefox | 19.3% | 48.9% | 31.8% |
| Google Chrome | 33.6% | 32.5% | 33.9% |
| Other | 18.9% | 45.9% | 35.2% |
| **Total** | 27.4% | 38.7% | 33.9% |

$\chi^2=16.298 \cdot df=4 \cdot \varphi=0.181 \cdot p=0.003$

**Table 6: Participants who read the Firefox UI content and endorsed Chrome, Firefox, or Neither as faster. Overall, 38.7% (p=.003) of participants endorsed Firefox as fastest, providing partial support for the idea that priming messages related to UI increase the perception of performance.**

In contrast to the source treatment condition, which had no significant or substantive effects in this model, the content treatment significantly increased the likelihood that participants would rate Firefox as faster. Reading an article about the improved performance of Firefox resulted in an 18% increase in likelihood (β=0.18, p=<.001) of endorsing Firefox as faster, whereas the UI treatment was less potent and less significant (β=0.08, p=.011), essentially rejecting the idea that respondents will think Firefox is faster after reading about UI improvements while confirming that respondents will think Firefox is faster after reading about performance.

As with our earlier investigation, we sought to determine if brand preferences were more powerful than priming. If it is the case that the priming effect is stronger than brand preferences, we should observe a greater likelihood of endorsing Firefox as faster given the performance prime, regardless of an individual's brand preference. More pointedly, some brands may be more resilient to priming effects than others.

|  | **Browser preference** | | **Source treatment** | | **Content treatment** | | **Full model** | |
| --- | --- | --- | --- | --- | --- | --- | --- | --- |
|  | *Likelihood* | *p* | *Likelihood* | *p* | *Likelihood* | *p* | *Likelihood* | *p* |
| *(Intercept)* | 0.52 | <.001 | 0.40 | <.001 | 0.31 | <.001 | 0.44 | <.001 |
| Prefers Google Chrome | -0.18 | <.001 |  |  |  |  | -0.19 | <.001 |
| Prefers other | -0.08 | .034 |  |  |  |  | -0.09 | .019 |
| Tech-focused press |  |  | -0.01 | .781 |  |  | -0.00 | .997 |
| Performance focus |  |  |  |  | 0.18 | <.001 | 0.18 | <.001 |
| UI focus |  |  |  |  | 0.08 | .011 | 0.08 | .007 |
| AIC | 2082.222 | | 2113.086 | | 2081.931 | | 2051.907 | |
| $X^2_{deviance}$ | p=.020 | | p=.892 | | p=.020 | | p=.006 | |

Table 7: Model predicting endorsement of Firefox as faster against browser preference (with Firefox preference as the base condition), source treatment, and content treatment.

To test this hypothesis, we fit a final logit model with an interaction term for content treatment against browser preference. As shown in Table 8, we see more evidence that priming effects may be stronger than brand preferences, though there are differences between brands and most of the results are only significant at the α=0.1 level. First, those participants who preferred Chrome were 14% more likely to endorse Firefox as faster after getting the performance treatment (β=0.14, p=.08), whereas they were not significantly more likely to endorse Firefox as faster given the UI treatment (β=0.11, p=.18). This is in contrast with the respondents who preferred alternative browsers, who were equally likely to rate Firefox as faster given either the performance (β=0.18, p=.05) or the UI treatment (β=0.18, p=.057). Thus, it appears that while priming does appear to be stronger than brand preferences, priming may be more robust against weaker brands.

To confirm that participants did not assign different levels of credibility to the two sources, which could have interacted with the efficacy of treatment, we referred to the questionnaire items that related to self-reported individual article credibility ratings. First, we recoded each article rating to a scale of [-3, +3] where +3 = 'Strongly agree' and -3 = 'Strongly disagree'. Next, we constructed a generalized article credibility score by summing all of the recoded article ratings. This score was then rescaled to lie between [0, 1], such that a score closer to 1 suggests high article credibility. The mean of the credibility score was .68 with an SD of .158. We then examined the mean credibility ratings over all conditions and found that credibility assessments were insignificantly different from each other. This affirms the results from the logit models examining the effect of source on likelihood to endorse Firefox as faster. It appears that for a general audience, there is no effect of source credibility on perception of performance regardless of which content participants were exposed to.

|  | **Endorsed Mozilla Firefox as Faster** | |
| --- | --- | --- |
|  | *Likelihood* | *p* |
| *(Intercept)* | 0.51 | <.001 |
| Performance content | 0.06 | .432 |
| UI content | -0.02 | .759 |
| Prefers Google Chrome | -0.27 | <.001 |
| Prefers alternate browser | -0.21 | .001 |
| Performance * Prefers Google Chrome | 0.14 | .081 |
| UI * Prefers Google Chrome | 0.11 | .184 |
| Performance * alternate browser | 0.18 | .050 |
| UI * alternate browser | 0.18 | .057 |
| AIC | 2052.417 | |
| $X^2_{deviance}$ | p=.026 | |

Table 8: Model predicting endorsement of Firefox as faster with an interaction term for content treatment against browser preference.

## DISCUSSION

With actual performance being about the same for various web browsers, finding ways to improve users' perceived performance is integral to optimizing browser quality. Our work has demonstrated the importance of media coverage and the resulting priming effect on users' perception of performance for web browsers.

As expected given Google's prevalence in the advertising sector [38], coverage in the mainstream media [33], and its primary marketing message that Chrome is "a faster browser," [12, 29], it is not surprising that Chrome is seen as fastest by default. However, when users are primed with an article about Mozilla Firefox, they were more likely to endorse it as faster than participants in the control group. This finding suggests that engaging with the media to present software as high-performing is an important step in influencing users' perceptions of that browser. More specifically, priming users about software's performance is an integral piece of overall perceived performance that must be considered by developers, and it also presents a potential opportunity for smaller companies to compete with large companies that have an established brand.

We also find that this effect is not simply an example of agenda setting, where the quantity of coverage is related to its prevalence in users' minds. The effect observed was indeed priming, as the message detailing Firefox UI updates was not as effective as the message about Firefox performance improvements in changing users' perceptions of performance. Specifically telling participants that Firefox was faster made participants perceive that Firefox was faster. This suggests that in order to be most effective in altering users' perceptions of a system, it is important to carefully curate the messages that they receive in a way that reinforces the desired takeaway message.

Additionally, the source of a story does not appear to alter the priming effect, and, for a general audience, it is equally important to engage with both technology and non-technology media sources to influence users' perceptions of performance. Our primary suggestion is that designers and developers work closely with marketing teams and the media to produce and widely distribute content regarding the performance of their technologies in order to boost user perceived performance.

**Limitations and Future Work**
One substantial limitation of our work is its short-term nature. Our results demonstrate that immediate perceptions can change based on priming, but previous work [27] has illustrated that beliefs changed as a result of perceiving the source of a message as credible are unstable and do not typically persist over time. Without longitudinal data, we cannot draw any conclusions regarding the persistence of the priming effect or whether repeated exposure to priming messages in the media affects respondents' perception of browser performance. Future work should examine this with regard to time to see whether a longer-term effect exists or how users' perceived performance of a technology product might change as an effect of longitudinal repeated messages about performance.

Next, while the general news source (USA Today) and tech-oriented source (The Verge) each resulted in a priming effect, this study represents a partial exploration of potential media sources. Additional work could explore the effect of a variety of news sources, such as those with different levels of popularity, recognition, and trustworthiness. Similarly, while we did ask respondents to rate the credibility of each article, we did not gather ratings of the technological credibility of the news source (USA Today, The Verge) as a whole. Future work could benefit from exploring how the perceived technological credibility of a news source might alter the priming effect observed in our study.

Our study was also performed with respondents residing in the United States with a desktop browser context, and the results might not generalize to other populations and use contexts, such as mobile. It could be beneficial to study the priming effect on perceived performance among different groups and cultures that may perceive time, brands, and/or the media differently [31] and in different use contexts.

The main goal of the present study was to investigate the priming effect for a less widely-used web browser. Alternatively, we did not explore the case of priming respondents with articles about Google Chrome, the most popular web browser [41], which could potentially produce different results. However, given the previously-discussed prevalence of Google's coverage in the mainstream media [33] and its marketing message that Chrome is "a faster browser" [12, 29], we could argue that Google is already harnessing the priming effect to its advantage, as illustrated by our finding that Chrome was seen as fastest by default. Our work demonstrates the importance of such priming, and we hope that other companies will also begin using it to improve perceived performance of their software.

Finally, for future exploration, our observations that priming did influence participants' perception of performance indicate an interesting direction for future work: namely, the assessment of perception of performance is not limited to psychophysical measures such as load times, response times, and user tolerance levels. Rather, our study demonstrates that user perceived performance can be influenced by message priming in the mass media. Future studies might explore the potential relationship between traditional psychophysical perceived performance measurements and the influence of messaging and priming on users' overall perception of performance.

**CONCLUSION**
With technical performance being similar for many systems, investigating ways of improving user perceived performance is especially important for designing and developing successful products. In our study, we primed users to influence the perceived performance of web browsers through an online survey. We found that reading priming messages about Mozilla Firefox improved participants' perceived performance of Firefox over Google Chrome, suggesting a potential opportunity for smaller companies to compete against large companies with an established brand. Further, we illustrated the benefits of priming with particular curated content over more general agenda setting, as the message regarding performance improvements resulted in more endorsement of Firefox as faster than the message about UI updates. Our results illustrate the importance of specific media messages in affecting users' perceptions of technology and suggest that designers and developers must consider the presentation of their software in the media when trying to improve perceived performance.

**ACKNOWLEDGMENTS**

We would like to thank our participants for their time and effort and our colleagues for their guidance and support.